\documentclass[journal]{IEEEtran}
\usepackage{float}
\usepackage{stfloats}
\usepackage{physics}
\usepackage{graphicx}
\usepackage{amsmath}
\usepackage{slashbox}
\usepackage{footnote}
\usepackage{threeparttable}
\usepackage{tabu,multirow}
\usepackage{amsthm, amssymb, amsfonts}
\usepackage{cite}
\ifCLASSINFOpdf
\else
\fi
\begin{document}

\title{SHEAR-net: An End-to-End Deep Learning Approach for Single Push Ultrasound Shear Wave Elasticity Imaging}

\author{
Tamim Ahmed and Md. Kamrul Hasan\IEEEauthorrefmark{1}

\thanks{T. Ahmed and M. K. Hasan are with the Department of Electrical and Electronic Engineering, Bangladesh University of Engineering and Technology, Dhaka-1205, Bangladesh. (e-mail: \IEEEauthorrefmark{1}khasan@eee.buet.ac.bd)}
\thanks{``This work has been submitted to the IEEE for possible publication. Copyright may be transfered without notice, after which this version may no longer ne accessible."}
}

\maketitle

\begin{abstract}
Ultrasound Shear Wave Elastography (USWE) with conventional B-mode imaging demonstrates better performance in lesion segmentation and classification problems. In this article, we propose SHEAR-net, an end-to-end deep neural network, to reconstruct USWE images from tracked tissue displacement data at different time instants induced by a single acoustic radiation force (ARF) with 100\% or 50\% of the energy in conventional use. The SHEAR-net consists of a localizer called the S-net to first localize the lesion location and then uses recurrent layers to extract temporal correlations from wave patterns using different time frames, and finally, with an estimator, it reconstructs the shear modulus image from the concatenated outputs of S-net and recurrent layers. The network is trained with 800 simulation and a limited number of CIRS tissue mimicking phantom data and is optimized using a multi-task learning loss function where the tasks are: inclusion localization and modulus estimation. The efficacy of the proposed SHEAR-net is extensively evaluated both qualitatively and quantitatively on 125 test set of motion data obtained from simulation and CIRS phantoms. We show that the proposed approach consistently outperforms the current state-of-the-art method and achieves overall 4--5 dB improvement in PSNR  and  SNR. In addition, an average gain of 0.15 in DSC and SSIM values indicate that the SHEAR-net has a better inclusion coverage area and structural similarity of the two approaches. The proposed real-time deep learning based technique can accurately estimate shear modulus for a minimum tissue displacement of 0.5$\mu$m and image multiple inclusions with a single push ARF.
\end{abstract}


\IEEEpeerreviewmaketitle

\section{Introduction}

\IEEEPARstart{U}{ltrasound} elastography (USE), a non-invasive clinical diagnostic technique, measures the mechanical property, the stiffness, that has a significant correlation with the tissue pathology. Adjunct to conventional B-mode ultrasound, USE improves the diagnostic quality and can provide important qualitative and quantitative information about tissue stiffness that may be helpful in many clinical applications, e.g., investigating diseases like fibrosis, cirrhosis, and hepatitis in the liver, diagnosing cancer in organs like breast and prostate \cite{ferraioli2012accuracy,chang2011clinical,barr2017wfumb}. In addition to extracting mechanical properties of tissues, USE has been successfully used to determine the state of muscles and tendons \cite{hatta2015quantitative}, assess stiffness in the brain and remove blood clotting \cite{ghanouni2015transcranial}.

The USE imaging work-flow starts with stimulation for tissue displacement, followed by an acquisition of echo signals usually known as ultrasound Radio Frequency (RF) signal and finally processing the RF data to reconstruct the elastography image. To generate tissue displacement, different excitation approaches, e.g., manual force, ARF, and vibration are used in clinically applicable USE techniques and among these techniques Quasi-static Elastography (QSE), Acoustic Radiation Force Impulse Imaging (ARFI), Shear Wave Elastography (SWE), Supersonic Shear Imaging (SSI), and Transient Elastography (TE) are vastly applied for cancer diagnosis and clinical management \cite{barr2017wfumb,nightingale2001feasibility,berg2012shear,bercoff2004supersonic,sandrin2003transient}. Several studies report that QSE with B-mode ultrasound improves the diagnosis and evaluation of breast lesions \cite{tan2008improving}. However, QSE has limitations as it is highly operator dependent and incapable of deeper organ imaging \cite{li2017quality}. Moreover, the quantitative information provided by SWE gives better performance compared to QSE \cite{yang2017qualitative}. Therefore, SWE has emerged as a new imaging tool that has high reproducibility, capable of deeper organ imaging and has low operator dependency compared to QSE \cite{sim2015value}. It has shown potential performance in breast, liver, and prostate lesion detection and diagnosis \cite{sang2017accuracy,yang2017qualitative,li2017quality}.

In SWE imaging, an automated stimulation of tissue displacement by ARF is induced to generate a shear wave propagation. As a result of this propagation, tissues are displaced in the normal direction of wave propagation. The initial challenge is to track such small tissue displacement over time. In most of the SWE algorithms, tissue displacement is estimated by using normalized cross-correlation between the tracked ultrasound reference and displaced echo data \cite{pinton2005real}. Ultrasound tracking, however, suffers from jitter, transducer bandwidth, Signal-to-Noise Ratio (SNR), kernel length, the correlation coefficient between RF-lines being tracked, and magnitude of tracked RF lines correlated to the tracking frequency \cite{cespedes1995theoretical,hollender2015single}. When shear wave speed (SWS) is estimated from such noisy tracked tissue motion, it is prone to erroneous estimation. Therefore, denoising schemes, e.g., particle filter, directional filter, and EMD-based denoising, are adopted in different reported articles to make the motion estimation robust \cite{deffieux2011effects,8169096}.
Two types of approaches are reported for estimating SWS from the denoised motion data. The first category of approaches called the time of flight (ToF) algorithms, locally estimate wave arrival time using the maximum displacement peaks \cite{rouze2012parameters,amador2017improved} or cross-correlation of time signals \cite{bercoff2004supersonic, song2012comb}. Although ToF-based algorithms are fast and implementable in real-time, they are not noise-robust because of noise amplification during inversion operation and misplacement of peaks \cite{wang2013precision}. The second category of methods for shear wave velocity estimation involves the frequency domain. These approaches use phase velocity estimated from the local maximum wave number, and two dimensional Fourier transformation on the time-space signal to estimate the phase velocity \cite{8485657,bernal2011material}. In both categories of approaches, the number of ARF pushes make a difference in the quality of the reconstructed images as described in LPVI and CSUE \cite{8485657,song2012comb}. Though current state-of-the-art is the LPVI technique, the efficacy of this algorithm largely depends on the window selection like other conventional approaches. Moreover, multiple pushes that may be required for improved SWE imaging will create the risk of tissue heating.

In recent years, Deep Neural Network (DNN) based methods have outperformed conventional state-of-the-art algorithms in signal and image processing tasks. DNN has made it possible to automatically detect a metastatic brain tumor and diagnose liver fibrosis and cardiac diseases \cite{7426413,8051114}. Also, profound imaging quality and accuracy have been achieved in MRI image reconstruction \cite{8067520}, classification and segmentation problem \cite{8051114}, and image denoising \cite{8340157} with the incorporation of DNN. In ultrasound elastography, DNN based classification and QSE image reconstruction \cite{wu2018direct} have been published. It is reported that DNN-based QSE image reconstruction algorithms can effectively extract, represent, and integrate highly semantic features without manual intervention \cite{wu2018direct}. Therefore, the DNN-based SWE image reconstruction algorithm can be an alternative to the existing conventional algorithms.

In this work, we propose SHEAR-net, a DNN-based noise robust high quality SWE image reconstruction method employing tracked tissue motion data induced by a single ARF pulse. The attributes of this proposed work are: 
\begin{itemize}
\item A novel architecture called the S-net which is a combination of 3-D CNN, Convolutional LSTM, and 2-D CNN. The S-net solves the inclusion localization problem  and reconstructs sharp inclusion boundary using the reflected wave patterns from tissue boundaries. The temporal correlation among these patterns are generated using recurrent layers;
\item A shear modulus estimation block using dense layers with skip connections that takes concatenated feature maps of the S-net and recurrent blocks as input;
\item Dynamic training of the S-net, recurrent layers and modulus estimation block with multi-task learning (MTL) loss function. The latter makes the SHEAR-net an end-to-end learning DNN approach;
\item The proposed technique can retain almost the same image quality with half of the ARF intensity required for the conventional algorithms and estimate shear modulus with greater accuracy for tissue displacement $\geq$0.5 $\mu$m;
\item A larger ROI for imaging to visualize multiple inclusion with just a single push;
\end{itemize}
The test results reveal that SHEAR-net, the first ever DNN-based SWE imaging technique can outperform the state-of-the-art algorithms both in quality and reconstruction time.
\section{Materials and method}
\begin{figure}[!t]
\centering
\includegraphics[width=\linewidth]{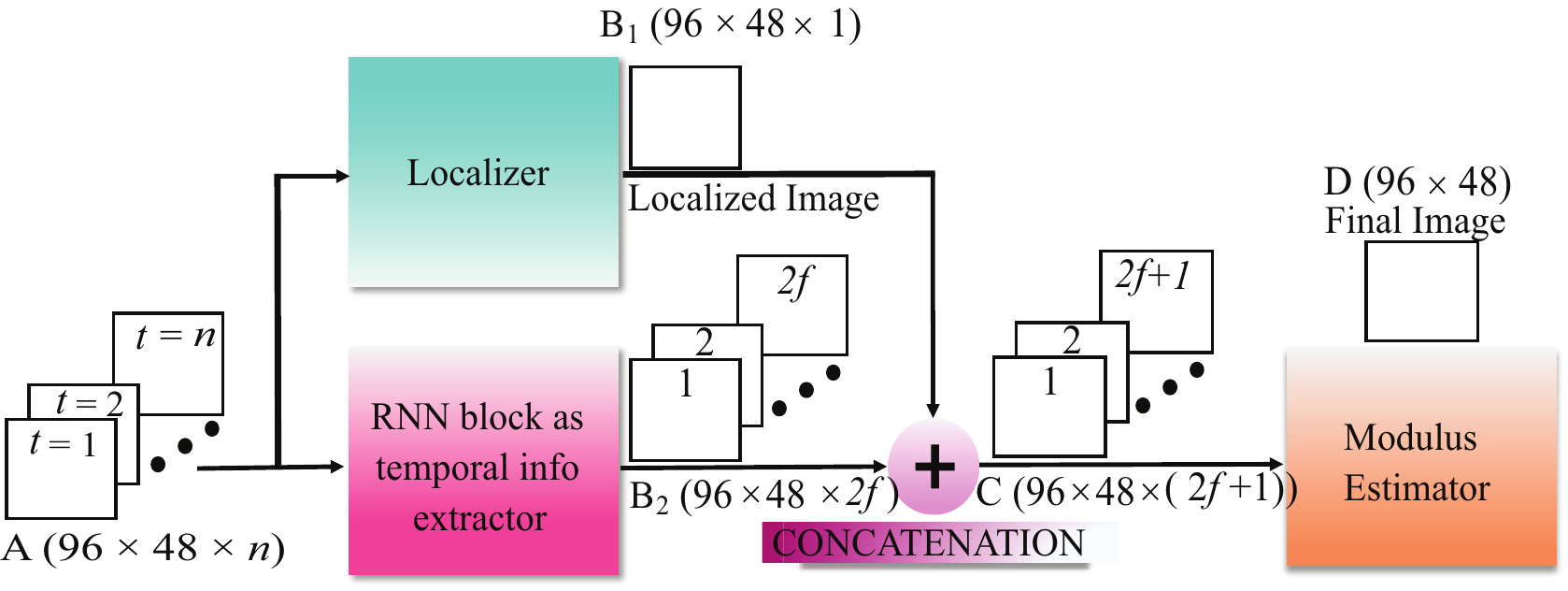}
\caption{The detailed block diagram of the proposed SHEAR-net.}
\label{pipe}
\end{figure}

In this section the newly proposed deep learning based approach, SHEAR-net, is adopted for shear modulus (SM) estimation from a single push ARF induced tissue displacement. First, we present the network architecture in Sec. II-A and its representative block diagram is shown in Fig. \ref{pipe}. We also explain its functionality by dividing the SHEAR-net into sub-blocks. The blocks are optimized by a multi-task learning loss function as described in Sec. II-B. The first task for our problem is to localize the inclusion position from the raw displacement data and for that, we propose the S-net. The displacement data is passed into the RNN-Block (RB) as illustrated in Fig. \ref{pipe} and temporal correlation among the time frames is extracted without changing the image dimension. The output of the RB is concatenated with the output of S-net and finally, the Modulus Estimator (ME) as shown in Fig. \ref{pipe} calculates the absolute modulus for each pixel. 
\subsection{Proposed SHEAR-net}
\subsubsection{S-net}
\begin{figure*}[!t]
\centering
\includegraphics[width=6.5 in]{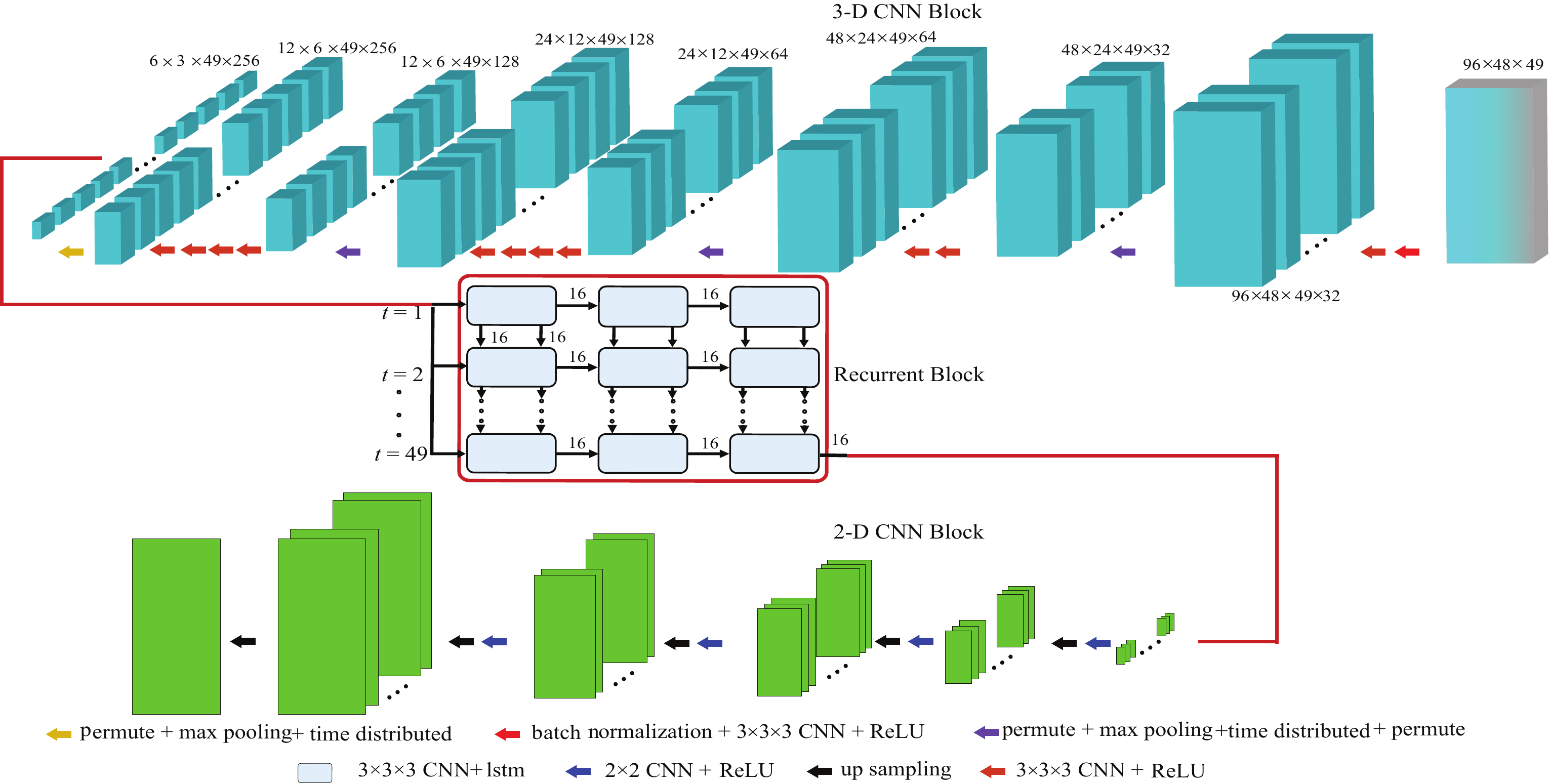}
\caption{The detailed feature map diagram of the proposed S-net. It is a 3-D CNN-recurrent block-2-D CNN combination.}
\label{re}
\end{figure*}

We have designed an S-net, a combination of 3-D CNN block, recurrent block, and 2-D CNN block as shown in Fig. \ref{re} to achieve sharper edges in the inclusion boundary and localize the inclusion by reconstructing a binary mask. The raw displacement data $\mathbf{D}$, is given as
\begin{equation}
\mathbf{D}=[\mathbf{D}_1~\mathbf{D}_2\cdot\cdot\cdot\mathbf{D}_{T_d}],~\mathbf{D}_t\in\mathbb{R}^{h\times w\times1}|~t=1,~2,\cdot\cdot\cdot,T_d
\label{eqn1}
\end{equation}
where $\mathbf{D}_t$ denotes the 3-D tissue displacement data with spatial dimension of $h \times w \times1$ at $t$ time frame and $T_d$ is the total frame count. The S-net first extracts a low level of spatio-temporal features $\mathbf{X}_{l_{st}}$ as the following
\begin{equation}
\begin{split}
\mathbf{X}_{l_{st}}&=[\mathbf{X}^1_{l_{st}}~\mathbf{X}^2_{l_{st}}\cdot\cdot\cdot\mathbf{X}^{T_d}_{l_{st}}],\\
\mathbf{X}^t_{l_{st}}&=\mathcal{F}_3(\mathbf{D_t};\theta)\in\mathbb{R}^{\frac{h}{m}\times \frac{w}{m}\times f}|~t=1,~2,\cdot\cdot\cdot,T_d
\end{split}
\end{equation}
where $\mathbf{X}^t_{l_{st}}$ denotes the encoded spatio-temporal feature maps with spatial dimension of $\frac{h}{m} \times \frac{w}{m} \times f$ at the $t$ time frame and the $\mathcal{F}_3(\cdot)$, $m$, $f$, and $\theta$ represent 3-D CNN operation, shrink coefficient on the spatial domain, feature map number, and network parameter, respectively. The feature maps are then passed into the recurrent block that has multiple ConvLSTM units as shown in Fig. \ref{re1} (c). Although the LSTM is a powerful tool to handle temporal correlation in a given sequence, for more general solutions of spatio-temporal forecasting problems, convolutional LSTM has the superiority in state transitions and holding spatial information. As illustrated in Fig. \ref{re1} (c), the ConvLSTM 2-D block takes in the current input $X_t$ with previous cell states $C_{t-1}$ and hidden states $H_{t-1}$ to generate the current cell state $C_t$ and hidden state $H_t$. The relation between inputs and gates are governed by
\begin{align}
i_t &=\sigma(W_{xi}*X_t+W_{hi}*H_{t-1}+b_i),\nonumber \\
f_t &=\sigma(W_{xf}*X_t+W_{hf}*H_{t-1}+b_f),\nonumber \\
C_t &=f_t\circ C_{t-1}+i_t\circ tanh(W_{xc}*X_t+W_{hc}*H_{t-1}+b_c),\nonumber \\
o_t &=\sigma(W_{xo}*X_t+W_{ho}*H_{t-1}+b_o),\nonumber \\
h_t &=f_t \circ tanh(C_t),\label{eqn1}
\end{align}
where $*$ and $\circ$ are defined as convolutional and element-wise matrix multiplication operator, respectively; $W_{xi}$, $W_{hi}$, $W_{xf}$, $W_{hf}$, $W_{xo}$, $W_{ho}$, $b_i$, $b_f$, and $b_o$ are convolutional parameters for the network and $i_t$, $f_t$, and $o_t$ represent input, forget and output gates, respectively. In the network, `$tanh$' is used as the activation function and $\sigma$ represents sigmoid operation. 
For a general input $\mathbf{X}=[\mathbf{X}_1\cdot\cdot\cdot\mathbf{X}_{T_d}]$, the S-net uses a ConvLSTM block for each time frame, $t$ and extracts hidden states, $\mathbf{H}$ and a high level of spatio-temporal features, $\mathbf{O}^t_{H_{st}}$ for each time frame as in the following
\begin{align}
\mathbf{O}_{H_{st}}&=[\mathbf{O}^1_{H_{st}}\mathbf{O}^2_{H_{st}}\cdot\cdot\cdot\mathbf{O}^{T_d}_{H_{st}}],\mathbf{R}=\mathcal{M}(\mathbf{O}_{H_{st}};\theta) \\
\mathbf{O}^t_{H_{st}}&=\mathcal{M}(\mathbf{X}_t;\theta)\in\mathbb{R}^{\frac{h}{m}\times \frac{w}{m}\times f}|~t=1,~2,\cdot\cdot\cdot,T_d\\
\mathbf{H}&=\mathcal{M}(\mathbf{X};\theta)\in\mathbb{R}^{\frac{h}{m}\times \frac{w}{m}\times f}
\label{eqn3}
\end{align}
where $\mathcal{M}(\cdot)$ indicates convolutional LSTM operation. A series of convolutional blocks extract deeper high level of spatio-temporal features using (5). And the final ConvLSTM layer shrinks the temporal length to 1 and outputs 3-D feature maps, $\mathbf{R}$ in (4). After the 4-D to 3-D feature map conversion using the recurrent block, the S-net uses consecutive 2-D CNN and upsampling layers to extract a high level of spatial features and restore the original image dimension by
\begin{equation}
\mathbf{X}_{H_{s}}=\mathcal{U}(\mathcal{F}_2(\mathbf{R};\theta))\in\mathbb{R}^{h\times w\times f},
\end{equation}
where $\mathcal{U}(\cdot)$ and $\mathcal{F}_2(\cdot)$ represent upsampling and 2-D CNN operation, respectively. Finally, the binary mask is reconstructed as
\begin{equation}
\mathbf{M}=\mathcal{F}_2(\mathbf{X}_{H_{s}};\theta)\in\mathbb{R}^{h\times w\times 1},
\end{equation}
which is our target feature extracted by the S-net that localizes the inclusion and in this convolutional layer, we have used `sigmoid' activation to keep the output within 0 to 1.
\subsubsection{RME Block}
\begin{figure}[!ht]
\centering
\includegraphics[width=3 in]{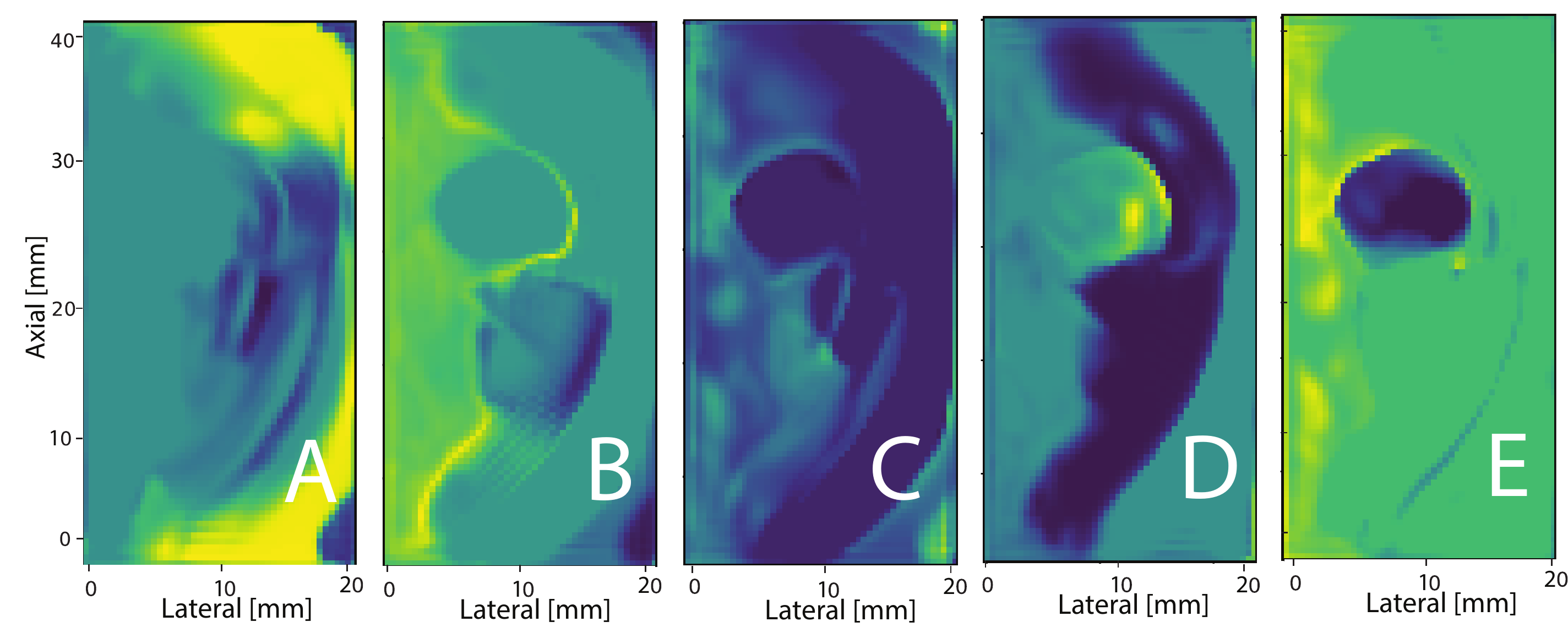}
\caption{Illustrating outputs of the recurrent layer that extracts temporal correlation from time frames and finds features related to the inclusion boundary as seen from the shades.}
\label{ref}
\end{figure}

The RME block consists of recurrent layers with skip connections and a modulus estimator for the reconstruction of the SM image. The recurrent layers, in this case, take tissue displacement data, $\mathbf{D}$ as input and learn different reflection patterns corresponding to wave propagation such as reflected waves from inclusion boundaries and from tissue boundaries as shown in Fig. \ref{ref}. The two recurrent layers use the tissue motion data and extracts the reflection patterns by (4) and (5). A skip connection between the hidden states of the first layer calculated form (6) and output of the second layer from (4) ensures smooth temporal feature propagation and is given by
\begin{equation}
\mathbf{R}^\tau=\mathbf{cat}(\mathbf{R},\mathbf{H})\in\mathbb{R}^{h\times w\times 2f},
\label{eqn9}
\end{equation}
where $\mathbf{cat}(\cdot)$ denotes the concatenation operation along the feature map axis. Some of the feature maps from \eqref{eqn9} are demonstrated in Fig. \ref{ref} and these reflected wave patterns are important for the SHEAR-net to inherently learn stiffness variation. These high level spatio-temporal feature maps are concatenated with the binary mask of S-net by
\begin{equation}
\mathbf{O}^\tau=\mathbf{cat}(\mathbf{R}^\tau,\mathbf{M})\in\mathbb{R}^{h\times w\times 2f+1},
\end{equation}
\begin{figure}[!t]
\centering
\includegraphics[width=\linewidth]{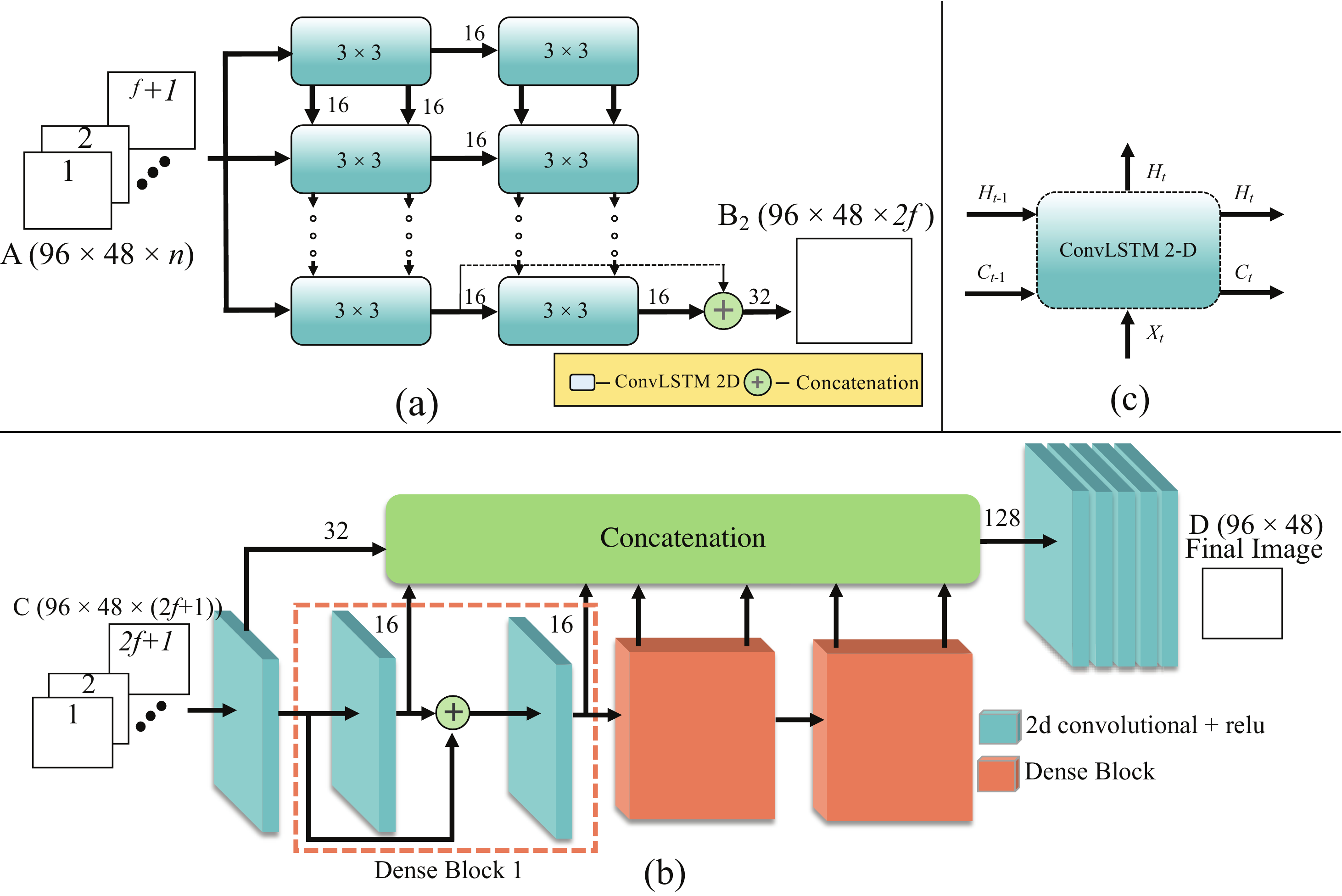}
\caption{Block diagram of the RME block consisting of (a) and (b); (a) recurrent layers of ConvLSTM and (b) ME, and (c) a single ConvLSTM block.}
\label{re1}
\end{figure}
The final block, ME as shown in Fig. \ref{re1} (b) estimates the shear modulus from the concatenated feature maps of $\mathbf{O}^\tau$. This feature map concatenation directs the ME to look at the reflection patterns from inclusion boundaries in the positions with high activation estimated from the S-net. The ME includes a convolutional layer and 3 dense blocks with skip connections. As seen in Fig. \ref{re1} (b), each dense block with 2 outputs, $\mathbf{E}_{ij}$, given by
\begin{equation}
\begin{split}
 \mathbf{E}_{i1}&=\mathcal{F}_2(\mathbf{O}^\tau;\theta)\in\mathbb{R}^{h\times w\times f}\\ 
\mathbf{E}_{i2}&=\mathcal{F}_2(\textbf{cat}(\mathbf{E}_{i1},\textbf{O}^\tau);\theta)\in\mathbb{R}^{h\times w\times f}
\end{split}
\end{equation}
are concatenated together, and finally the 2-D CNN layers reconstruct the SM image from the concatenated feature maps as
\begin{equation}
\mathbf{P}=\mathcal{F}_2(\textbf{cat}(\mathbf{E}_{11},~\mathbf{E}_{12},\cdot\cdot\cdot,\mathbf{E}_{i2});\theta)\in\mathbb{R}^{h\times w\times 1}.
\end{equation}
Each input feature maps in (18) has different receptive field and therefore, capture high level of spatial features for accurate SM estimation. The skip connections in the dense layers allows this smooth flow of features to concatenate with the forward layers and help reduce the gradient vanishing problem. 
\subsection{Multi-Task Learning (MTL) Loss Function}
The proposed SHEAR-net optimizes its weight based on a novel MTL loss function. The first task of the S-net is to localize the inclusion boundary and the second task is to estimate the shear modulus value of each pixel using RB with ME (RME). However, optimizing one task is not independent of the other as the output of S-net is concatenated with the output of RB as an input of ME.
For SWE imaging, we have already discussed the efficacy of S-net. The goal is to label each pixel of the image either as an object (inclusion in our case) or as background. In most cases, a major portion of the ROI belongs to the background class. As the ratio of the inclusion area to the background area is very small, the network has low accuracy in estimating the shear modulus inside the inclusion compared to that of the background in most cases. Therefore, we have adopted the IoU loss function to direct the SHEAR-net towards the inclusion area and emphasize on the overlap of the ground truth and the predicted region. If $P$ and $G$ are the set of predicted and ground truth binary labels, respectively, then the IoU function (also known as the Jaccard similarity index), is defined as
\begin{equation}
 J_c(P,G) = \dfrac{|P \cap G|}{|P \cup G|} = \dfrac{|P \cap G|}{|P - G| + |P \cap G| + |G - P|}.
\end{equation}
The IoU loss is defined as
\begin{equation}
    \label{eqn_iou_loss}
 \text{IoU}= 1 - Jc(P, G) = 1 - \dfrac{|P \cap G|}{|P \cup G| + \epsilon},
\end{equation}
where $\epsilon$ is a small safety factor added in order to handle division by zero, $1e-7$ for our case.
For optimizing the RME, we have defined a loss function called Modulus Loss, $L_m$ for the purpose of estimating the shear modulus. The modulus loss is given by
\begin{equation}
L_m=\sum_{j=1}^{m}(G_j-P_j)^2,
\label{mod}
\end{equation}
where $G_j$ and  $P_j$ denote ground truth and SHEAR-net predicted pixels, respectively, and $m$ is the number of pixels present in the image. Now, our target is to optimize both the loss functions at the same time.  To this end, the joint loss function for the proposed SHEAR-net is given by
\begin{table}[t]
\centering
\caption{\label{param}Simulation Paramters For Shear Wave Generation}
\begin{tabu}{|c|c|}
\hline
\textbf{Parameter} & \textbf{Value}\\
\hline
ARF intensity, A   &  $1\times10^6$~N/m$^3$ \\\hline
$\sigma_x$, $\sigma_y$, $\sigma_z$   & 0.21 mm, 0.21 mm, 0.43 mm  \\\hline
Focusing point & $x_0$, $y_0$, $z_0$\\\hline
\multirow{2}{*}{\textbf{Medium}} & Nearly incompressible\\\cline{2-2}
& linear, isotropic, elastic solid\\\hline
Poison's ratio, $\nu$ & 0.499\\\hline
Density, $\rho$ & 1000~kg/m$^3$\\\hline
Time for ARF excitation & 200 $\mu$s\\\hline
Time for shear wave propagation & 18 ms\\\hline
FEM size & $40\times20\times40$~mm\\\hline
Inclusion radius & 2.5~mm\\\hline
Inclusion co-ordinate & 8~mm, 0~mm, 20~mm\\\hline
Mesh element and size & tetrahedral and 0.2 mm\\\hline
\end{tabu}
\end{table}
\begin{equation}
 J=\alpha L_m+(1-\alpha)\text{IoU}
\label{weight}
\end{equation}
For our work, the value of the weight factor, $\alpha$ is selected to be 0.5 after observing the consistency in the results for a different range of values.
\begin{table}[b]
\centering
\caption{\label{var}Variable Parameters For Simulation Phantom Generation}
\begin{tabular}{|c|c|}
\hline
\textbf{Parameter} & \textbf{Value}\\
\hline
ARF intensity, A & $1\times10^6,  2\times10^6$~N/m$^3$\\\hline
Inclusion radius &  random number within 1--5 mm\\\hline
Inclusion co-ordinate & randomly generated\\\hline
Background stiffness (BS) & 10, 20~kPa\\\hline
Inclusion (Sphere Oval) modulus& 2, 4, 6, 8, 10 times of BS\\\hline
\end{tabular}
\end{table}
\section{Experimental Setup}
\subsection{Simulation of Shear Wave Propagation}
In order to generate a shear wave (SW) in an elastic medium (see Table \ref{param}), we need to apply ARF. It is reported that ARF modeled with Gaussian distribution \cite{palmeri2017guidelines} is highly correlated with the impulse response generated by a transducer. We have used COMSOL Multiphysics 5.1 to simulate SW propagation and further processing was done in MATLAB (Mathworks, Natick, MA) software to generate 2-D tissue displacement. In our simulation, ARF was modeled as a Gaussian impulse given by
\begin{equation}
\centering
    \text{ARF}=Aexp(-(\frac{(x-x_0)^2}{2\sigma_x^2}+\frac{(y-y_0)^2}{2\sigma_y^2}+\frac{(z-z_0)^2}{2\sigma_z^2})),
\end{equation}
where $x_0$, $y_0$, $z_0$ represent ARF focusing point, $\sigma_x$, $\sigma_y$, $\sigma_z$ define ARF beam width in $x$, $y$, $z$ direction, respectively. For safety issues, ARF intensity $A$ is chosen to keep the maximum displacement around 20$\mu$m to mimic the displacement required for real life tissue imaging. The parameters used for the simulation of shear wave propagation are given in Table \ref{param}.
\subsection{Simulation Dataset}
First, we import the simulation data (i.e., gold standard and training data) in MATLAB using the COMSOL-MATLAB interface for ultrasound tracking. In Field II, we have designed an L12-4 probe and tracked 2-D tissue displacement data using \cite{palmeri2006ultrasonic} over a time span of 8 ms. From the tracked motion data we have extracted 2-D displacement data of size $96\times48$ for 49 time frames. This size is taken because of memory constraint. The gold standard for the training of the network is the shear modulus image that is generated in COMSOL Multiphysics 5.1 and processed in MATLAB. Our goal is to create a dataset that has varied samples of breast and liver phantoms with different tissue stiffness, different inclusion shape, size, and position. For this reason, we have simulated a variety of finite element models that can mimic human breast (female) and liver tissues. These models particularly mimicked breast fibroadenoma and homogeneous liver tissue.  Table \ref{var} presents the parameters for varied simulation data. For this initial study, we have generated data for a homogeneous inclusion in a homogeneous background. The stiffness values were taken so as to obtain the previously reported models \cite{johns1987x}. The ARF intensity is varied so that a maximum tissue displacement of 20$\mu$m or 10$\mu$m is obtained. The ARF excitation point is kept fixed at the center for all the models. However, the center of the inclusions is not aligned with the center of focus in most of the data. This brings position variation in the dataset along with the variation of tissue stiffness. The total number of samples in our dataset is 800. The number of samples for spherical inclusion, fibroadenoma, and liver mimicking tissue is 300, 300 and 200, respectively.
\subsection{CIRS Phantom Dataset}
For this study, we have downloaded CIRS experimental phantom (Model 049A, CIRS Inc., Norfolk, VA, USA) data from ftp://ftp.mayo.edu received from Ultrasound Research Laboratory, Department of Radiology, Mayo Clinic, USA. From the provided data we have used Type III and Type IV phantoms for our study having a background stiffness of 25 kPa each and inclusion stiffness of 45 kPa and 80 kPa, respectively. Both types of phantoms had 4 different inclusion sizes, i.e., 2 mm, 4 mm, 6 mm, and 10 mm. The phantom has a sound speed of 1540 m/s, ultrasound attenuation of 0.5 dB/cm/MHz and the inclusions are centered around 30 mm and 60 mm from the phantom surface. In this study, the ARF pulse is focused at 30 mm with a duration of 400 $\mu$s and the push frequency is 4.09 MHz. The push beam is generated by 32 active elements shifted by 16 elements from the end of the L7-4 probe and placed on each side of the inclusion. A single push acquisition is used in our study and acquired data is processed using the auto-correlation algorithm to get the motion data with a frame rate of 11.765 kHz and spatial resolution of 0.154 mm. All the CIRS phantom data are pre-processed using a 15 point locally weighted smoothing window \cite{cleveland1981lowess} as tissue displacements are affected by high-frequency ultrasound tracking noise also known as jitter.
\subsection{Training, Optimization}
Our model is implemented using Keras library backend with Tensrflow. We have split the dataset into training, validation, and test sets with 380, 160, and 121 simulation phantoms, 49 time frames each, respectively. We have also split our limited CIRS phantom data in the same way with 8 in training, 4 in validation and 4 in the test set. With end-to-end learning, we have trained the full SHEAR-net from scratch. Normalized 2-D tissue displacement data for 49 time frames are directly used as input for training without augmentation. Any augmentation that may misplace the displacement patterns will slow down the convergence rate. We have used a batch size of 16 because of memory constraint and ADAM as the optimizer with the initial learning rate of $5\times10^{-3}$. The learning rate is varied with cosine annealing and converges at around 100 epochs. As for the training labels, 2 sets of labels are generated: label 1 for the S-net and label 2 for the RME. Label 1 is a binary mask having a pixel value of 1 inside the inclusion and 0 outside. Label 2 is the true modulus image with the absolute shear modulus of each pixel. For each label, the SHEAR-net optimizes the loss function in \eqref{weight} and outputs the predicted modulus image. Note that for all the 49 time frames of a sample data, we have used the above 2 labels. Given the input sequence and the target image, the loss function is calculated using the forward propagation and the parameters of the network are updated using the backpropagation. This is repeated for a number of iterations that is 120 for our case.
\subsection{Evaluation metrices}
We evaluate our proposed method's performance in the test set by computing signal-to-noise ratio (SNR) as defined in \cite{wu2018direct}, peak-signal-to-noise-ratio (PSNR) and structural similarity index (SSIM) as defined in \cite{schlemper2018deep}, and S{\o}rensen--Dice coefficient (DSC) ($ = {(2|P \cap G|)}/({|P| + |G|})$)  as quantitative evaluation indices. We have also computed the runtime of our proposed method using a PC with GPU NVIDIA GeForce GTX 1080 Ti and CPU Intel Core i7-7700K @4.20GHz.
\begin{table}[t]
\centering
\caption{\label{homo_1}Quantitative Performance Evaluation Of The Reconstructed SM Image Using The LPVI Technique And The Proposed SHEAR-net For Two Homogeneous Phantoms}
\begin{tabular}{|c|c|c|c|c|}
\hline
 & \multicolumn{2}{c|}{\textbf{Type I}} & \multicolumn{2}{c|}{\textbf{Type II}} \\ \cline{2-5}
\multirow{-2}{*}{ \textbf{Indices}}&  \textbf{LPVI} &  \textbf{SHEAR-net} &  \textbf{LPVI} &  \textbf{SHEAR-net} \\ \hline
\textbf{PSNR{[}dB{]}} & 15.23 & 22.52 & 16.23 & 20.98 \\ \hline
\textbf{SNR{[}dB{]}} & 23.11 & 39.41 & 24.56 & 35.82 \\ \hline
\textbf{SSIM} & 0.45 & 0.94 & 0.63 & 0.87 \\ \hline
\begin{tabular}[c]{@{}c@{}}\textbf{Background}\\ \textbf{mean$\pm{}$SD [kPa]}\end{tabular} & \begin{tabular}[c]{@{}c@{}}8.191\\ $\pm{1.773}$\end{tabular} & \begin{tabular}[c]{@{}c@{}}6.816\\ $\pm{0.199}$\end{tabular} & \begin{tabular}[c]{@{}c@{}}3.521\\ $\pm{0.926}$\end{tabular} & \begin{tabular}[c]{@{}c@{}}3.282\\ $\pm{0.112}$\end{tabular} \\ \hline
\end{tabular}
\end{table}
\begin{figure}[!t]
\centering
\includegraphics[width=3 in]{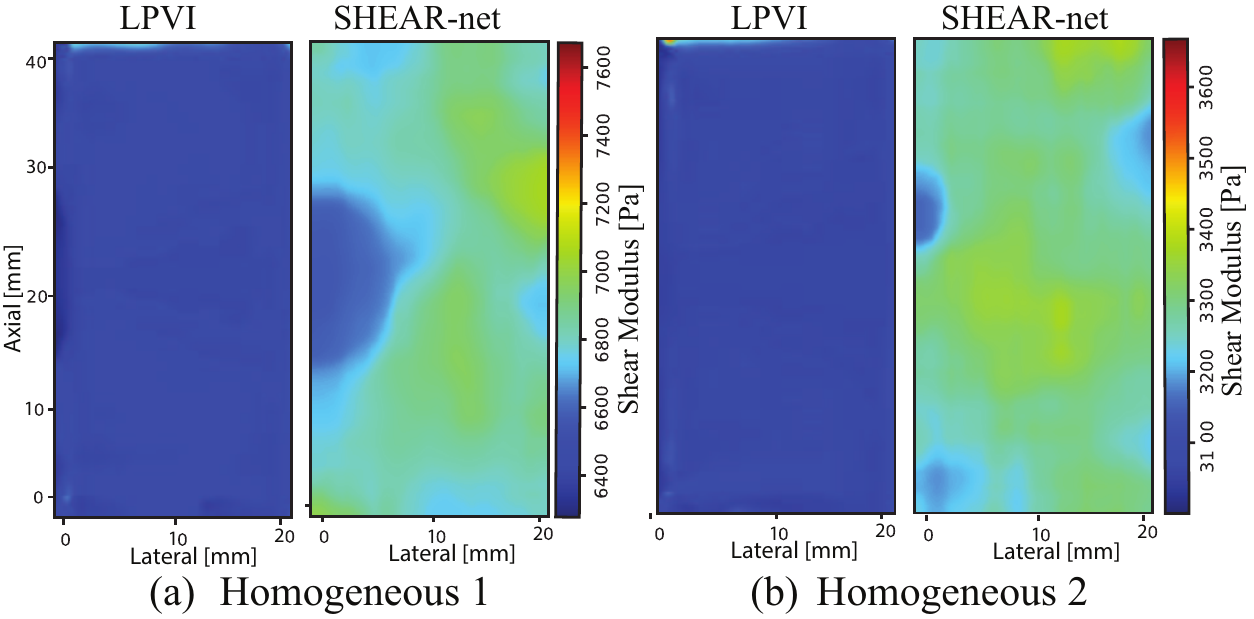}
\caption{Reconstructed 2-D SM image of two types of homogeneous simulation phantom using the LPVI technique and the proposed SHEAR-net.}
\label{homo_2}
\end{figure}
\section{Results}
In this section, we present the results of our proposed SHEAR-net for both the simulation and experimental phantom data and also compare its performance with the most recent state-of-the-art algorithm: local phase velocity imaging (LPVI) \cite{8485657}.
\subsection{Simulation Study}
Our simulation test set contains homogeneous phantoms mimicking liver tissue and phantoms with spherical and oval-shaped inclusion that mimic breast tissue with fibroadenoma. We first show the results on homogeneous phantoms. Figure \ref{homo_2} presents 2-D SM images reconstructed using the LPVI method and our proposed SHEAR-net. The true mean SM for Type I is 6.663 kPa and that for Type II is 3.335 kPa. The reconstructed SM images of these two phantoms by the SHEAR-net show more homogeneity in comparison to that for the LPVI algorithm. The estimated values of  mean SM$\pm{}$standard deviation (SD) as presented in Table \ref{homo_1} are evidence of this fact. Other quantitative indices, i.e,. PSNR, SNR, SSIM, and DSC presented in Table \ref{homo_1} indicate that SHEAR-net has the ability to reconstruct significantly better quality 2-D SM image compared to that of the LPVI technique. 
\begin{table}[!ht]
\centering
\caption{Comparative Results Between LPVI And SHEAR-net For Inclusions Of Different Shape And Modulus }
\label{tmi_1}
\begin{tabular}{| c|c|c|c|c|}
\hline
  & \multicolumn{2}{c|}{ {\begin{tabular}[c]{@{}c@{}}\textbf{Type I (Spherical} \\ \textbf{inclusion)}\end{tabular}}} & \multicolumn{2}{c|}{ {\begin{tabular}[c]{@{}c@{}}\textbf{Type II (Oval}\\ \textbf{inclusion)}\end{tabular}}} \\ \cline{2-5}
\multirow{-2}{*}{ \textbf{Index}}&  \textbf{LPVI} &  \textbf{SHEAR-net} & \textbf{LPVI} &  \textbf{SHEAR-net} \\ \hline
\textbf{PSNR{[}dB{]}} & 18.06 & 25.01 & 17.11 & 28.44 \\ \hline
\textbf{SNR{[}dB{]}} & 19.12 & 35.98 & 18.21 & 33.57 \\ \hline
\textbf{SSIM} & 0.77 & 0.90 & 0.55 & 0.97 \\ \hline
\textbf{DSC} & 0.59 & 0.78 & 0.62 & 0.81 \\ \hline
\begin{tabular}[c]{@{}c@{}}\textbf{Background} \\ \textbf{mean$\pm{}$SD [kPa]}\end{tabular} & \begin{tabular}[c]{@{}c@{}}5.163\\ $\pm{0.107}$\end{tabular} & \begin{tabular}[c]{@{}c@{}}3.438\\ $\pm{0.025}$\end{tabular} & \begin{tabular}[c]{@{}c@{}}11.306\\ $\pm{0.106}$\end{tabular} & \begin{tabular}[c]{@{}c@{}}3.393\\ $\pm{0.025}$\end{tabular} \\ \hline
\begin{tabular}[c]{@{}c@{}}\textbf{Inclusion}\\ \textbf{mean$\pm{}$SD [kPa]}\end{tabular} & \begin{tabular}[c]{@{}c@{}}14.236\\ $\pm{0.060}$\end{tabular} & \begin{tabular}[c]{@{}c@{}}15.822\\ $\pm{0.014}$\end{tabular} & \begin{tabular}[c]{@{}c@{}}25.584\\ $\pm{0.070}$\end{tabular} & \begin{tabular}[c]{@{}c@{}}23.455\\ $\pm{0.016}$\end{tabular} \\ \hline
\end{tabular}
\end{table}
\begin{figure}[!t]
\centering
\includegraphics[width=3 in]{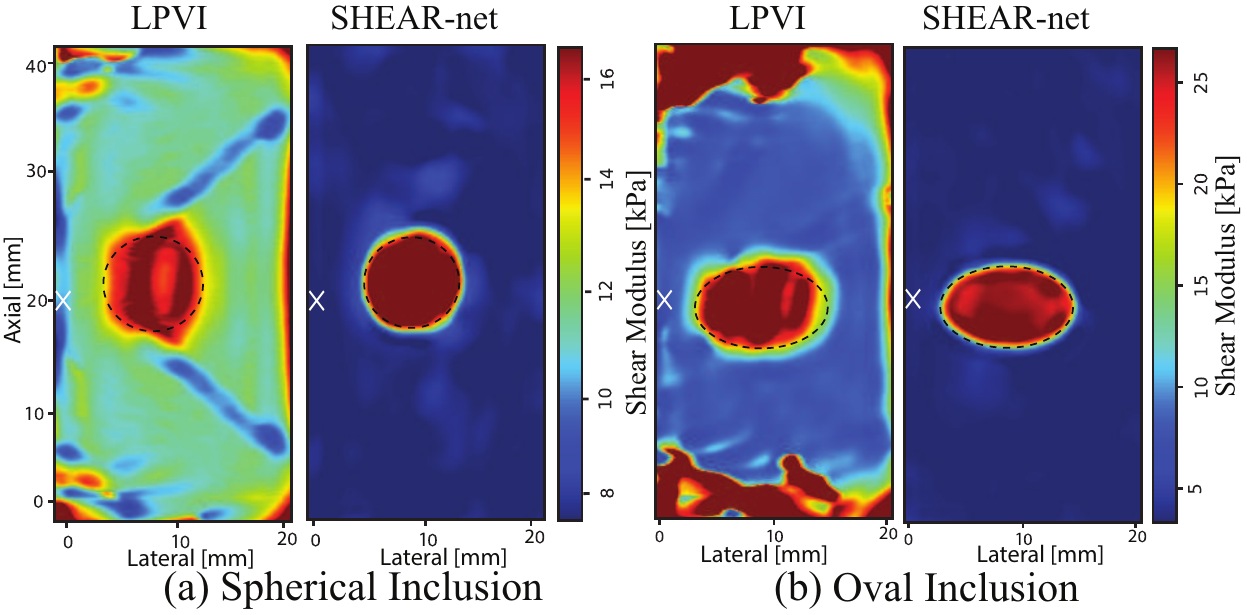}
\caption{ Qualitative comparison of the reconstructed SM image from the tracked motion data of simulation phantoms with inclusion. White cross marks indicate the focusing point of the ARF.}
\label{tmi_2}
\end{figure}

Next, we present the results on simulation data with inclusion. Our simulation dataset contains two shapes of inclusion: Type I- spherical inclusion, and Type II- oval inclusion. Figure \ref{tmi_2} presents the 2-D reconstructed SM images using the LPVI method and our proposed SHEAR-net and Table \ref{tmi_1} shows numerical indices evaluated for these reconstructed images. The spherical inclusion sample has a mean SM of 3.33 kPa for the background and 16.587 kPa for the inclusion. Oval inclusion sample has a mean SM of 3.33 kPa for the background and 23.499 kPa for the inclusion. The illustrations and mean$\pm{}$SD both indicate that both the techniques have a good inclusion coverage area. However, the reconstructed SM images by the LPVI technique has more background noise and a little contrast variation inside the inclusion.  On the contrary, the reconstructed images by the SHEAR-net demonstrate more overall homogeneity and less noise in the background and thus we get high PSNR and SNR values. These images have a more accurate mean with small SD both inside and outside the inclusion. Moreover, SHEAR-net reconstructs a sharper boundary around the inclusion irrespective of the shape and has higher structural similarity compared to that of the LPVI technique which is evident from the DSC and SSIM values.
\begin{table}[t]
\centering
\caption{\label{force_1}Quantitative Comparison Between LPVI And The SHEAR-net For Experiment With Force Variation}
\begin{tabular}{| c|c|c|c|c|}
\hline
  & \multicolumn{2}{c|}{ {\begin{tabular}[c]{@{}c@{}}\textbf{Type I (100\% }\\ \textbf{force)}\end{tabular}}} & \multicolumn{2}{c|}{ {\begin{tabular}[c]{@{}c@{}}\textbf{Type II (50\%}\\ \textbf{force)}\end{tabular}}} \\ \cline{2-5} 
\multirow{-2}{*}{ \textbf{Index}}&  \textbf{LPVI} &  \textbf{SHEAR-net} &  \textbf{LPVI} &  \textbf{SHEAR-net} \\ \hline
\textbf{PSNR{[}dB{]}} & 20.34 & 25.01 & 18.06 & 23.39 \\ \hline
\textbf{SNR{[}dB{]}} & 22.52 & 35.98 & 19.12 & 30.66 \\ \hline
\textbf{SSIM} & 0.89 & 0.91 & 0.77 & 0.90 \\ \hline
\textbf{DSC} & 0.73 & 0.78 & 0.59 & 0.76 \\ \hline
\begin{tabular}[c]{@{}c@{}}\textbf{Background} \\ \textbf{mean$\pm{}$SD [kPa]}\end{tabular} & \begin{tabular}[c]{@{}c@{}}6.607\\ $\pm{0.107}$\end{tabular} & \begin{tabular}[c]{@{}c@{}}3.412\\ $\pm{0.025}$\end{tabular} & \begin{tabular}[c]{@{}c@{}}11.306\\ $\pm{0.107}$\end{tabular} & \begin{tabular}[c]{@{}c@{}}3.438\\ $\pm{0.025}$\end{tabular} \\ \hline
\begin{tabular}[c]{@{}c@{}}\textbf{Inclusion}\\ \textbf{mean$\pm{}$SD [kPa]}\end{tabular} & \begin{tabular}[c]{@{}c@{}}15.338\\ $\pm{0.060}$\end{tabular} & \begin{tabular}[c]{@{}c@{}}15.822\\ $\pm{0.014}$\end{tabular} & \begin{tabular}[c]{@{}c@{}}14.236\\ $\pm{0.060}$\end{tabular} & \begin{tabular}[c]{@{}c@{}}15.157\\ $\pm{0.014}$\end{tabular} \\ \hline
\end{tabular}
\end{table}
\begin{figure}[!t]
\centering
\includegraphics[width=3 in]{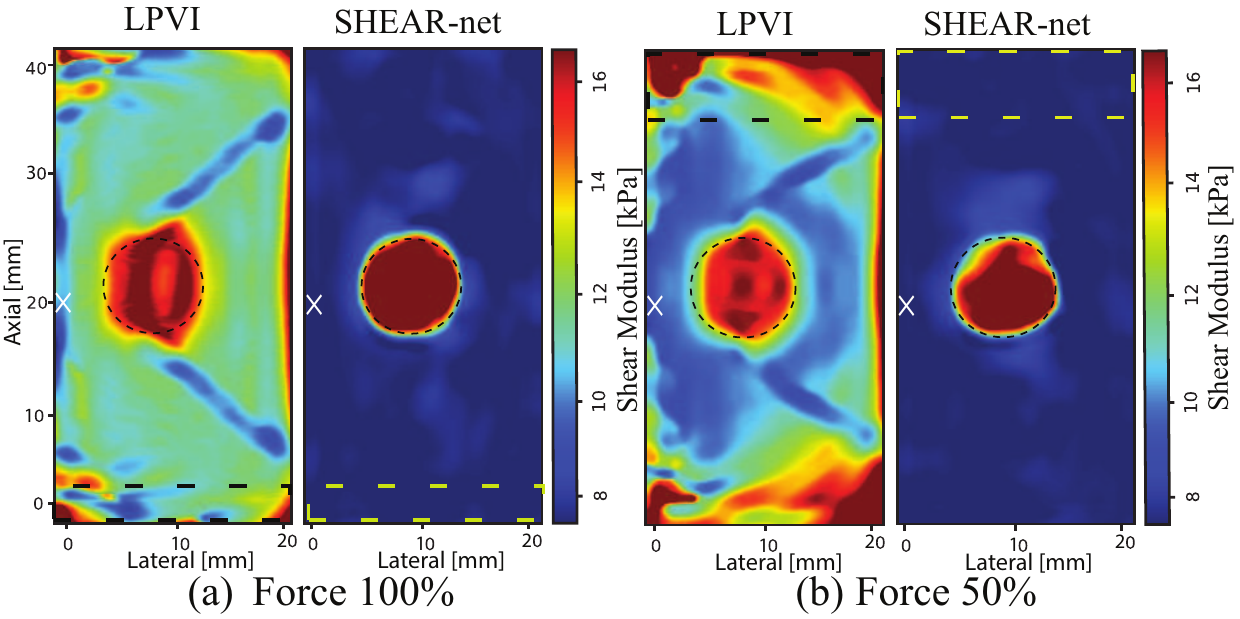}
\caption{Effect of ARF intensity variation on the reconstruction of SM image. Dashed boxes indicate regions where tissue displacement is $<1\mu m$.}
\label{force_2}
\end{figure}
\begin{table}[!t]
\centering
\caption{\label{size_1}Reconstruction Results For Experiment With Inclusion Size Variation}
\begin{tabular}{| c|c|c|c|c|}\hline
{\begin{tabular}[c]{@{}c@{}}\textbf{Inclusion}\\ \textbf{diameter}\end{tabular} }& \multicolumn{2}{c|}{ {\textbf{Type I (10 mm)}}}  & \multicolumn{2}{c|}{ {\textbf{Type II (4 mm)}}} \\ \hline
\textbf{Index}&  \textbf{LPVI} &  \textbf{SHRAR-net}  &  \textbf{LPVI} &  \textbf{SHEAR-net} \\ \hline
\textbf{PSNR{[}dB{]}} & 19.39 & 22.95 &  8.92 & 16.84 \\ \hline
\textbf{SNR{[}dB{]}} & 16.76 & 21.77 & 12.31 & 20.43 \\ \hline
\textbf{SSIM} & 0.78 & 0.90 &  0.49 & 0.87 \\ \hline
\textbf{DSC} & 0.65 & 0.71 & 0.49 & 0.87 \\ \hline
\begin{tabular}[c]{@{}c@{}}\textbf{Background}\\ \textbf{mean$\pm{}$SD [kPa]}\end{tabular} & \begin{tabular}[c]{@{}c@{}}17.005\\ $\pm{0.111}$\end{tabular} & \begin{tabular}[c]{@{}c@{}}6.707\\ $\pm{0.26}$\end{tabular} &  \begin{tabular}[c]{@{}c@{}}14.057\\ $\pm{0.165}$\end{tabular} & \begin{tabular}[c]{@{}c@{}}6.583\\ $\pm{0.25}$\end{tabular} \\ \hline
\begin{tabular}[c]{@{}c@{}}\textbf{Inclusion}\\ \textbf{mean$\pm{}$SD [kPa]}\end{tabular} & \begin{tabular}[c]{@{}c@{}}28.546\\ $\pm{0.065}$\end{tabular} & \begin{tabular}[c]{@{}c@{}}28.496\\ $\pm{0.015}$\end{tabular} &  \begin{tabular}[c]{@{}c@{}}15.864\\ $\pm{0.057}$\end{tabular} & \begin{tabular}[c]{@{}c@{}}9.125\\ $\pm{0.013}$\end{tabular} \\ \hline
\end{tabular}
\end{table}
\begin{figure}[!ht]
\centering
\includegraphics[width=3 in]{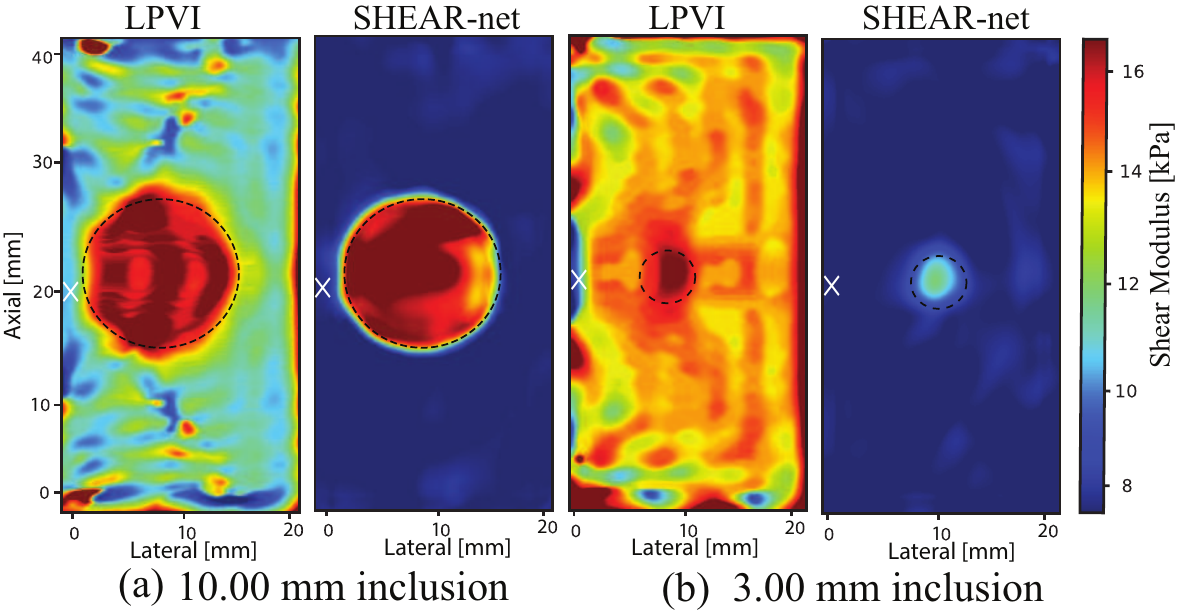}
\caption{SM reconstruction of inclusions of different sizes: (a) 10mm and  (b) 3mm using the LPVI technique and the proposed SHEAR-net.}
\label{size_2}
\end{figure}

Now, we evaluate the quality of reconstructed SM image to demonstrate the robustness of the technique against ARF intensity variation, inclusion size variation, and position variation. The effects of these variations on the reconstructed SM image will be discussed in the sequel. First, Fig. \ref{force_2} and Table \ref{force_1} show the results for ARF intensity variation. The phantom for this experiment has a mean SM of 3.337 kPa for background and 16.587 kPa for inclusion. We have induced two different ARF intensity: 100\% force refers to the intensity that can create 20 $\mu$m maximum tissue displacement and that 50\% force refers to 10 $\mu$m maximum tissue displacement. Reducing the force results in more background noise in the LPVI technique compared to our proposed SHEAR-net and is evident from the PSNR and SNR values presented in Table \ref{force_1}. Moreover, the change in the values of DSC, SSIM, and mean$\pm{}$SD both for the inside and outside of the inclusion for lowering the ARF intensity is more drastic for the LPVI technique compared to that for the SHEAR-net. Another important feature of SHEAR-net to notice in the reconstructed images is the dashed region with $0.5\mu m<d<1\mu m$, where $d$ indicates the tissue displacement. The mean values in the dashed region for the LPVI technique and SHEAR-net when the force is 100\% are 7.756 kPa and 3.523 kPa, respectively. When the force is halved, SHEAR-net retains almost the same mean value (i.e., 3.597 kPa) of SM inside the dashed region. However, the mean value (i.e., 13.256 kPa) of SM for the LPVI method has a very large deviation from the true SM value.
\begin{table}[!t]
\centering
\caption{\label{pos_1}Results To Evaluate Robustness Against Inclusion Position Variation}
\begin{tabular}{| c|c|c|c|c|}
\hline
 & \multicolumn{2}{c|}{  {\begin{tabular}[c]{@{}c@{}}\textbf{Type I (Position} \\ \textbf{top)}\end{tabular}}} & \multicolumn{2}{c|}{ {\begin{tabular}[c]{@{}c@{}} \textbf{Type II (Position}\\ \textbf{bottom)}\end{tabular}}} \\ \cline{2-5} 
 \multirow{-2}{*}{ \textbf{Index}}&   \textbf{LPVI} &   \textbf{SHEAR-net }&   \textbf{LPVI} &   \textbf{SHEAR-net} \\ \hline
\textbf{PSNR{[}dB{]}} & 18.67 & 29.11 & 16.61 & 28.41 \\ \hline
\textbf{SNR{[}dB{]}} & 15.33 & 30.89 & 13.22 & 31.29 \\ \hline
\textbf{SSIM} & 0.36 & 0.99 & 0.22 & 0.98 \\ \hline
\textbf{DSC} & 0.21 & 0.81 & 0.23 & 0.79 \\ \hline
\begin{tabular}[c]{@{}c@{}}\textbf{Background} \\ \textbf{mean$\pm{}$SD [kPa]}\end{tabular} & \begin{tabular}[c]{@{}c@{}}48.708\\ $\pm{0.100}$\end{tabular} & \begin{tabular}[c]{@{}c@{}}6.607\\ $\pm{0.024}$\end{tabular} & \begin{tabular}[c]{@{}c@{}}48.144\\ $\pm{0.106}$\end{tabular} & \begin{tabular}[c]{@{}c@{}}6.638\\ $\pm{0.025}$\end{tabular} \\ \hline
\begin{tabular}[c]{@{}c@{}}\textbf{Inclusion}\\ \textbf{mean$\pm{}$SD [kPa]}\end{tabular} & \begin{tabular}[c]{@{}c@{}}67.301\\ $\pm{0.132}$\end{tabular} & \begin{tabular}[c]{@{}c@{}}56.968\\ $\pm{0.021}$\end{tabular} & \begin{tabular}[c]{@{}c@{}}53.517\\ $\pm{0.025}$\end{tabular} & \begin{tabular}[c]{@{}c@{}}45.189\\ $\pm{0.005}$\end{tabular} \\ \hline
\end{tabular}
\end{table}

Next, we present the comparative results for varying inclusion size. In Fig. \ref{size_2}, we show the reconstructed images for inclusions of two different diameters. And Table \ref{size_1} demonstrates the quantitative indices obtained for them. Type I with 10 mm diameter inclusion has a mean SM of 6.691 kPa for background and 28.299 kPa for inclusion. Type II with 3 mm diameter inclusion has a mean SM of 6.671 kPa for background and 12.108 kPa for inclusion. Compared to the LPVI technique, the SHEAR-net shows greater insensitivity against the inclusion size variation. It is evident that the SHEAR-net is able to reconstruct SM images of small inclusion having a diameter of around 3 mm and also of the moderate size inclusion with a diameter of 10 mm. Although the LPVI technique can reconstruct SM images of relatively large inclusions, it shows below average performance for the small inclusions. The SHEAR-net is thus found to be more robust for inclusion size variation compared to LPVI. 

\begin{figure}[!t]
\centering
\includegraphics[width=3 in]{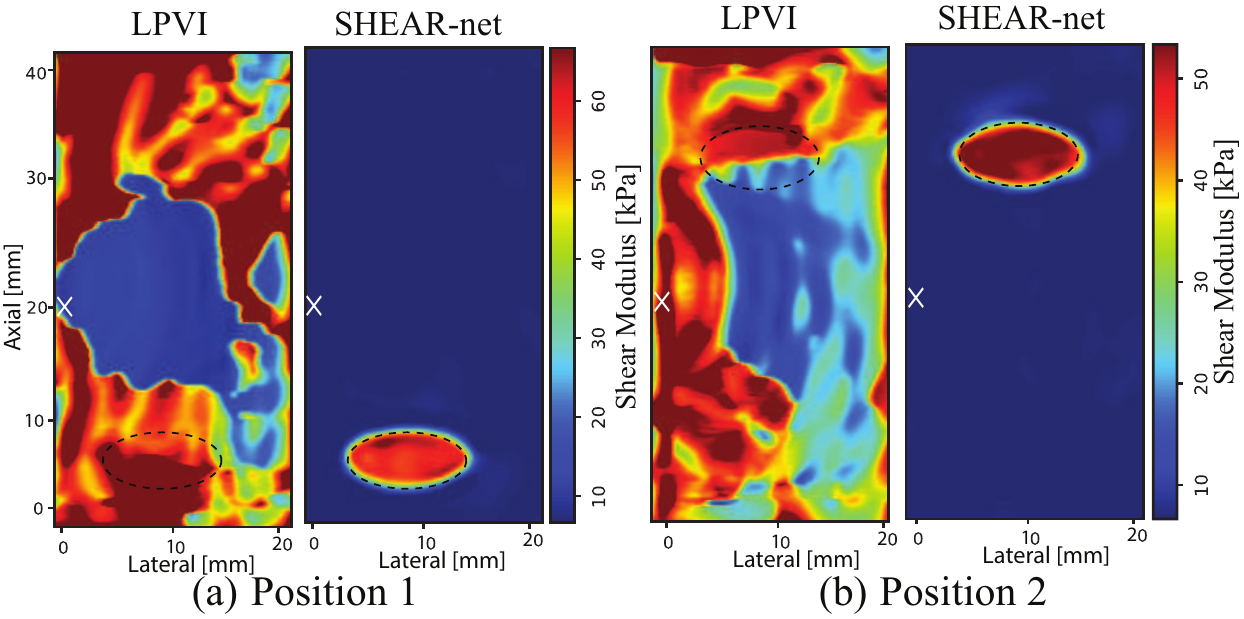}
\caption{Illustrating the robustness of SHEAR-net when the inclusions are centered 10-15 mm apart from the ARF focus point. The LPVI technique fails completely to reconstruct the lesions.}
\label{pos_2}
\end{figure}

Finally, last but not least important observation in our experiment is the inclusion position variation. To observe the robustness of the techniques in imaging inclusions that are positioned 10-15 mm far apart from the ARF focus point. We use Type I phantom that has a mean SM of 6.675 kPa for background and 57.271 kPa for inclusion and Type II phantom that has a mean SM of 6.674 kPa for background and 45.965 kPa for inclusion. The results of this observation are presented in Table \ref{pos_1}. A qualitative comparison is illustrated in Fig. \ref{pos_2}. Form the quantitative indices it is evident that the SHEAR-net is able to reconstruct high quality SM images that have inclusion center 10-15 mm apart from the ARF focus point. On the contrary, the LPVI technique suffers from the inability to reconstruct SM images in regions where the tissue displacement is below 1 $\mu$m as discussed earlier; it fails completely for the inclusions as shown in Fig. \ref{pos_2}.
\vspace{-0.5 cm}
\subsection{CIRS Phantom with Inclusion Study}
\begin{figure}[!t]
\centering
\includegraphics[width=3 in]{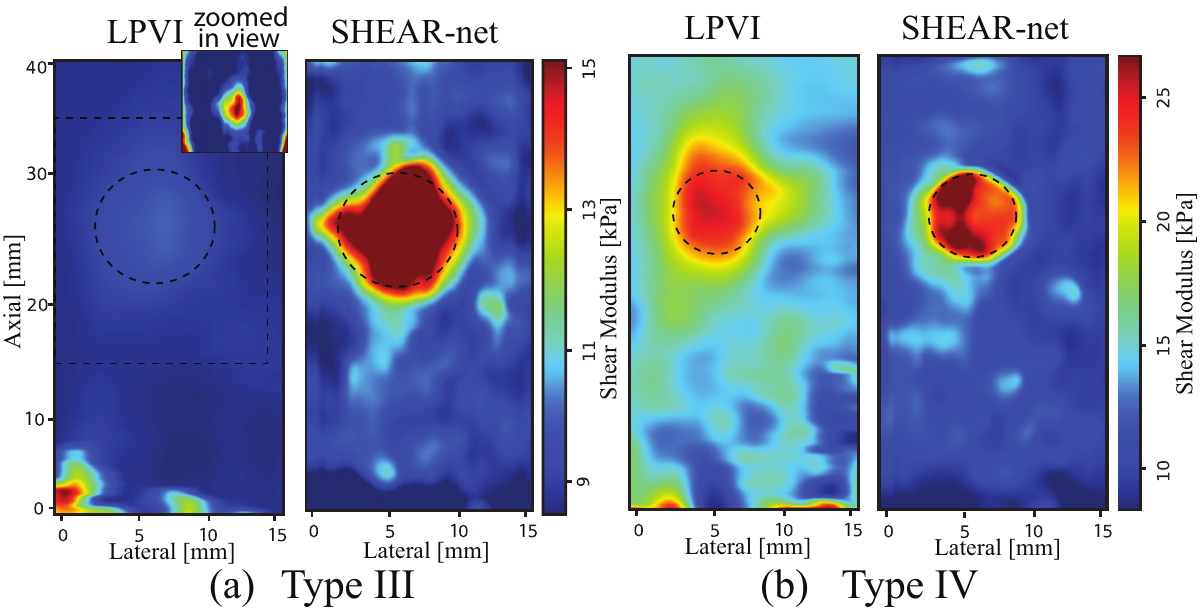}
\caption{2-D reconstructions of CIRS experimental phantom data: (a) Type III inclusion (b) Type IV inclusion; the zoomed-in-view of the marked region is illustrated to clearly visualize stiffness variation inside the lesion.}
\label{real_2}
\end{figure}

Figure \ref{real_2} demonstrates the 2-D reconstructed SM image of CIRS experimental phantom data with inclusion and Table \ref{real_1} presents the evaluated quantitative indices for these images. Each dataset has a mean SM value of 8.83 kPa in the background, and the inclusions of Type III and Type IV have mean SM values of 15.01 kPa and 24.68 kPa, respectively. We can observe contrast variation inside the inclusion from the zoomed-in-view of reconstructed images of Type III phantom using the LPVI technique. Also, the LPVI technique reconstructs a more noisy background and has little structural similarity for the Type IV inclusion. Therefore, the DSC and SSIM values are small compared to the proposed SHEAR-net. On the contrary, the homogeneity in the background and the better coverage of the inclusion in the reconstructed images by the proposed SHEAR-net are evident from the mean$\pm{}$SD values. Although the index values in Table \ref{real_1} for the Type III phantom are lower compared to that of the Type IV phantom for both the techniques, the performance of the LPVI technique declines more compared to the proposed SHEAR-net when the stiffness difference between the background and inclusion is small. The proposed SHEAR-net might be able to reconstruct better quality images if more CIRS experimental phantom data were available for training.
\begin{table}[!t]
\centering
\caption{\label{real_1}Results Of 2-D SM Reconstruction Using The LPVI And The SHEAR-net On 2 CIRS Phantom Data}
\begin{tabular}{|c|c|c|c|c|}
\hline
  & \multicolumn{2}{c|}{ \textbf{Type III}} & \multicolumn{2}{c|}{ {\color[HTML]{000000} \textbf{Type IV}}} \\ \cline{2-5} 
\multirow{-2}{*}{ \textbf{Index}} &  \textbf{LPVI} &  \textbf{SHEAR-net} &  \textbf{LPVI} &  \textbf{SHEAR-net} \\ \hline
\textbf{PSNR {[}dB{]}} & 10.62 & 14.22 & 20.78 & 21.61 \\ \hline
\textbf{SNR {[}dB{]}} & 11.21 & 15.31 & 23.5 & 25.80 \\ \hline
\textbf{SSIM} & 0.69 & 0.74 & 0.82 & 0.86 \\ \hline
\textbf{DSC} & 0.54 & 0.67 & 0.65 & 0.72 \\ \hline
\textbf{\begin{tabular}[c]{@{}c@{}}Background\\ mean$\pm{}$SD {[}kPa{]}\end{tabular}} & \begin{tabular}[c]{@{}c@{}}8.519\\ $\pm{}0.032$\end{tabular} & \begin{tabular}[c]{@{}c@{}}9.124\\ $\pm{}0.026$\end{tabular} & \begin{tabular}[c]{@{}c@{}}8.499\\ $\pm{}0.031$\end{tabular} & \begin{tabular}[c]{@{}c@{}}10.89\\ $\pm{}0.025$\end{tabular} \\ \hline
\textbf{\begin{tabular}[c]{@{}c@{}}Inclusion\\ mean$\pm{}$SD {[}kPa{]}\end{tabular}} & \begin{tabular}[c]{@{}c@{}}11.47\\ $\pm{}0.011$\end{tabular} & \begin{tabular}[c]{@{}c@{}}14.44\\ $\pm{}0.013$\end{tabular} & \begin{tabular}[c]{@{}c@{}}19.09\\ $\pm{}0.007$\end{tabular} & \begin{tabular}[c]{@{}c@{}}23.18\\ $\pm{}0.017$\end{tabular} \\ \hline
\end{tabular}
\end{table}

Lastly, we present the average results of 125 test cases in In Table \ref{res} including 121 simulation data and 4 CIRS phantom data. We test the robustness of the proposed technique for ARF variation and inclusion shape, size, stiffness, and position variation. The proposed SHEAR-net shows superior performance over the LPVI technique in terms of every index. It is evident from the PSNR and SNR values that the reconstructed images have a more homogeneous background with less noise compared to that of the LPVI technique. In addition, the average SSIM score of 0.94 and DSC score of 0.758 demonstrate the proposed technique's efficacy in reconstructing images with high structural similarity and a good coverage area of the inclusion, respectively. On the contrary, the performance of the LPVI technique for the test set shows that it is less robust to the variations mentioned earlier. Moreover, the reconstruction time for the SHEAR-net is 0.17 s and thus makes it currently the fastest SM image reconstruction algorithm to our knowledge. Finally, we have observed that multiple push could improve the performance of the LPVI technique, however, Table \ref{res} demonstrates that the proposed SHEAR-net can perform above that mark with just a single push. 
\begin{table}[!h]
\centering
\caption{\label{res}Quantitative Comparison Between LPVI And SHEAR-net On 125 Test Phantoms}
\begin{tabular}{| c|c|c|c|c|c|}\hline
Index & \begin{tabular}[c]{@{}l@{}}PSNR\\{[}dB]\end{tabular} & \begin{tabular}[c]{@{}l@{}}SNR\\{[}dB]\end{tabular} & SSIM & DSC & \begin{tabular}[c]{@{}l@{}}Run Time\\{~}{~}{~}(S)\end{tabular}  \\ 
\hline
\textbf{LPVI}  & 18.56 & 20.65&0.79&0.65&3712\\\hline
\textbf{SHEAR-net}  & 22.604&25.94&0.758&0.76&0.17\\\hline
\end{tabular}
\end{table}
\vspace{-0.5 cm}
\subsection{Discussion}
In this study, we present a new technique called the SHEAR-net for the shear modulus imaging in soft tissues. This is the first ever DNN-based SM image reconstruction algorithm from a single ARF pulse induced ultrasound 2-D tissue displacement data. The study shows promising results using the SHEAR-net in SM image reconstruction with high noise robustness, accurate shape representation, position independence and large 2-D ROI. The proposed technique can accurately estimate SM from a single ARF pulse induced tissue displacement data. Moreover, we have shown that the SHEAR-net is able to retain almost the same imaging quality even at half the ARF intensity level that is generally used in conventional imaging for displacement generation. We have used our algorithm to produce results on simulation and CIRS phantom data. Due to resource limitation, we could not study the efficacy of the SHEAR-net on \emph{in-vivo} data.\\
\begin{figure}[!t]
\centering
\includegraphics[width=3 in]{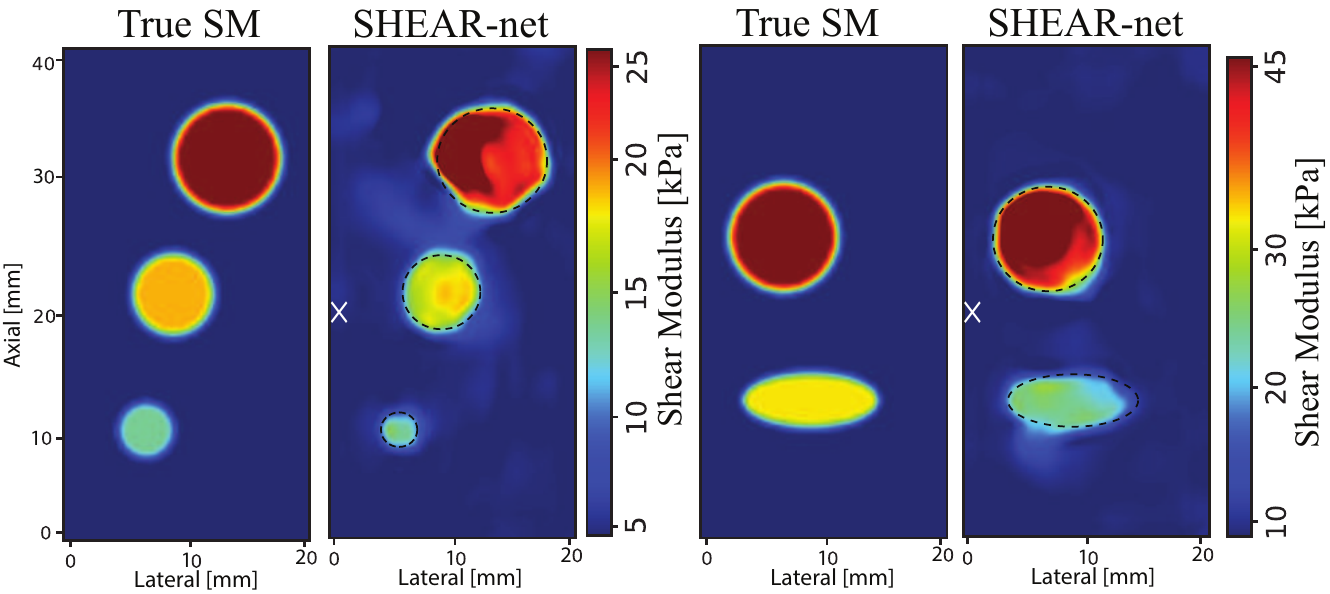}
\caption{One of the unique features of SHEAR-net. It is able to reconstruct multiple inclusion with a clear contrast variation to indicate different stiffness. (a) The gold standard of the simulation phantoms (b) Reconstructed SM images using SHEAR-net.}
\label{three}
\end{figure}
Now, we discuss the insight behind the parameters chosen for performance evaluation. Firstly, we look at the significance of the level of ARF intensity in SWE imaging. The intensity and tissue displacement are proportional to each other. High-intensity ARF creates high magnitude tissue displacement and the wave propagation covers a greater distance. This results in better quality SM image reconstruction as evident from the qualitative illustration in Fig. \ref{force_2} and quantitative evaluation in Table \ref{force_1}. However, high-intensity ARF excitation raises the risk associated with tissue heating \cite{liu2014thermal,doherty2013acoustic}. The level of ARF excitation, allowable clinically, generates tissue displacement of 20 $\mu$m \cite{liu2014thermal,doherty2013acoustic}. Therefore, we have set this value as 100$\%$ force in our simulation. We have observed that a higher force contributes to better quality SWE imaging but for tissue heating, it is not practically implementable. Our proposed SHEAR-net has the unique feature of SWE image reconstruction at 50$\%$ force maintaining almost the same quality as that of 100$\%$ and the same ROI dimension. This makes SWE imaging with SHEAR-net safer, more reliable and practical. Reducing the ARF intensity results in a smaller distance propagation of the induced shear waves. Our observations have shown that the best imaging area for the conventional approaches is an ROI where the tissue displacement is $\geq$1$\mu$m. We have illustrated in Fig. \ref{force_2} that as the tissue displacement decreases below 1$\mu$m, the quality of the reconstructed SM images gradually decreases. The quantitative values given in Table \ref{force_1} also support this observation. On the contrary, the SHEAR-net can reconstruct SM image for tissue displacement $\geq$0.5$\mu$m. This creates a promising opportunity to generate shear waves with lower ARF intensities and also maintain a larger ROI for SM image reconstruction. 

Note that with the conventional algorithms, the window for SWE imaging as seen in the modern ultrasound machines, e,g., Siemens Accuson S2000 is small. Multiple acquisitions and windows are required to visualize the whole region of 4$\times$4 cm$^2$ area, as seen in the B-mode image. With the proposed SHEAR-net, the observation window can be greater than the diagnostic windows in modern ultrasound machines making it possible to observe a 4$\times$2 cm$^2$ area in a single excitation. The immensely interesting and unique feature of the SHEAR-net as illustrated in Fig. \ref{three} is its ability to reconstruct SM images with multiple inclusions from the tracked tissue displacement data of a single push. For this demonstration, we have experimented with three sets of multiple inclusions phantom with different SM values for each inclusion. Note that these multiple-inclusion data were not included in our training and validation set. The LPVI technique fails completely to reconstruct an inclusion centered around 10-15 mm apart from the ARF focus point. From this observation, we can conclude that the SHEAR-net is the only existing technique that can reconstruct the SM images of multiple inclusions with a single push and maintain visually differentiable contrast for each stiffness. Moreover, the presence of different inclusions with shape and stiffness variation does not impact the reconstruction quality. Finally, the whole process is independent of the ARF focus point as all the results produced in this paper have the same focus point for ARF excitation irrespective of the position of the inclusion center.

The temporal resolution is an important factor for both the conventional approaches and the proposed SHEAR-net. For memory constraints, we have taken a sampling rate of 8 kHz. However, in the conventional approaches up to 12.5 kHz sampling frequency is taken to get high temporal resolution and better quality image reconstruction. We have observed that taking a high sampling rate increases the SWE image quality. Therefore, using 8 kHz sampling frequency in a constrained environment is one limitation of the proposed SHEAR-net. Another limitation is the lack of diversity in the data. For this initial study, our dataset includes random inclusion positions, sizes, and stiffness variations for two specific shapes. Therefore, our future investigation will include experimental phantoms and \emph{in-vivo} clinical ultrasound SWE data with diverse and more complex tissue structures. We have observed that the conventional algorithms perform better imaging with increased spatial resolution. However, we could not increase the spatial resolution for SHEAR-net because of the memory constraint. For our dataset, we have used images of the size 96$\times$48. Finally, due to the memory constraint our training batch size was 16. However, we have tested with sizes 14 and 15 too. Our observations showed that increasing the batch size also improved the performance of the SHEAR-net.
\section{Conclusion}
This paper has introduced SHEAR-net, a novel deep learning based shear modulus image reconstruction technique from single ARF pulse induced 2-D tissue displacement over a time period. The proposed architecture relies on a novel S-net to localize inclusion and RME block to estimate the SM values for each point. The SHEAR-net has demonstrated promising qualitative and quantitative performance both in simulation and CIRS phantom study. It has reconstructed SM images from tissue displacement generated by half of the ARF intensity generally used for the conventional algorithms. We have shown that the SHEAR-net can accurately estimate the SM for tissue displacement of $\geq$0.5$\mu$m and thus can maintain a larger ROI compared to the conventional algorithms. Moreover, the half ARF intensity level allows the SHEAR-net to perform imaging without any safety concern associated with tissue heating. 
In addition, the proposed technique can reconstruct multiple inclusions with contrast variation within the an ROI. Our algorithm has estimated shear modulus from the noisy tracked tissue displacements real-time. The memory constraints in spatial and temporal resolution presently are keeping the performance of the proposed SHEAR-net capped. The future work will focus on breaking these boundaries, and a detailed \emph{in-vivo} study.
\section*{Acknowledgment}
This work has been supported by HEQEP UGC (CP$\#$096/BUET/Win-2/ST(EEE)/2017), Bangladesh. We thank Dr. Matthew Urban, Ultrasound Research Laboratory, Department of Radiology, Mayo Clinic, USA for supporting us with the CIRS phantom data.
\ifCLASSOPTIONcaptionsoff
  \newpage
\fi
\bibliographystyle{IEEEtran}
\bibliography{reference}
\end{document}